\documentclass[11pt]{article}

\usepackage{amsmath}
\usepackage{amsfonts}

\newcommand{\dd}{\, \mathrm{d}}
\newcommand{\totder}[2]{\frac{\dd #1}{\dd #2}}

\newcommand{\pathdd}[1]{\left[\mathcal D #1 \right]}

\newcommand{\reals}{\mathbb{R}}

\newcommand{\Tr}{\mathrm{Tr}\, }
\usepackage{amsmath}
\usepackage{amsfonts}
\newcommand{\exval}[1]{\left\langle #1 \right\rangle}
\newcommand{\rnd}[1]{\left( #1 \right)}
\newcommand{\sqr}[1]{\left[ #1 \right]}
\newcommand{\crl}[1]{\left\{ #1 \right\}}

\usepackage{amsmath}
\usepackage{amsfonts}
\newlength{\slashoffset}

\newcommand{\ovslash}[1]{%
\settowidth{\slashoffset}{$#1\slash$}%
\addtolength{\slashoffset}{-1pt}%
#1\hspace{-0.5\slashoffset}\slash%
\hspace{0.5\slashoffset}%
\hspace{-5pt}%
}

\usepackage{hyperref}

\newcommand{\eps}{\varepsilon}
\title{abs}
\begin{document}

\begin{flushright}
  {\small
  MIT-CTP/4163}
\end{flushright}

\begin{center}

  {\huge Double-trace deformations, holography and the $c$-conjecture}
  \vspace{10mm}
  
  {\large \bf Andrea Allais}
  \vspace{3 mm}
  
  {\it Center for Theoretical Physics,
  
  Massachusetts Institute of Technology,
  
  Cambridge, MA 02139, USA}
\end{center}

\begin{abstract}
A double-trace deformation is the simplest perturbation of a conformal field theory that has a gravity dual. In this paper we review the existing results for the case in which the deformation is composed from a scalar operator, and extend them to the case of a spinor operator. In particular we check the validity of the $c$-conjecture along the RG flow induced by the deformation, using both Cardy's $c$-function and the recent proposal by Myers and Sinha of a $c$-function from entanglement entropy.
\end{abstract}

\section{Introduction}

An interesting way of perturbing a conformal field theory is to add the integral of a local operator to the conformal action. A particularly simple choice of perturbation, in the framework of the holographic correspondence, is to add the square of an operator $\mathcal O$ which is dual to a fundamental field in the gravity theory. 

This kind of perturbation is called double trace deformation and it has been extensively studied on both sides of the duality, at least in the case in which the operator is a scalar operator. It has been shown \cite{Witten:2001ua}\cite{Klebanov:1999tb}\cite{Hartman:2006dy}\cite{Berkooz:2002ug} that the dual operation of perturbing the conformal field theory is to appropriately change the  boundary conditions imposed on bulk fields. The two point function of the operator $\mathcal O$ in the perturbed field theory has been shown \cite{Klebanov:1999tb}\cite{Hartman:2006dy} to agree at leading order\footnote{The large $N$ limit and the associated factorization of correlation functions are crucial in simplifying the effects of the perturbation.} in $N$ on both sides of the duality, and the renormalization group flow induced by the perturbation has been given a geometric description on the gravity side \cite{Gubser:2002zh}. In fact, if the conformal dimension $\Delta$ of the operator $\mathcal O$ is less then $d/2$, the operator $\mathcal O^2$ is relevant\footnote{Because of the large $N$ limit, the dimensions of operators are additive, so that the dimension of $\mathcal O^2$ is $2\Delta$.}, and it triggers a renormalization group flow away from the conformal point, which can be shown to end on another conformal point, where the operator has dimension $d-\Delta > d/2$. It has also been possible to compute the difference in central charge between these two conformal field theories, and the same result has been obtained from both sides of the duality. The change turned out to be negative, in support of Cardy's $c$-conjecture \cite{Cardy:1988cwa}\cite{Gubser:2002zh}\cite{Gubser:2002vv}.

The aim of this paper is to extend these results to the case of a spinor field. In doing so we will find convenient to review some of the results obtained for the scalar field, and to discuss and clarify the way in which the perturbation is dualized as a change in the boundary conditions on the fields. 

The paper is organized as follows. In the next section we will discuss the issue of boundary conditions, and show how the two point function of the operator $\mathcal O$ can be obtained from both sides of the duality. We will review the known results on the scalar field and extend them to the spinor field. 

In the following section we will discuss the $c$-conjecture. We will review Cardy's proposal for the $c$-function as well as the proposal in \cite{Myers:2010xs} for constructing a $c$-function from the entanglement entropy that is meaningful in any (even and odd) dimensions. We will show how both kinds of $c$-functions can be computed from holography, and we will compute the difference in their value between the two conformal points connected by the RG flow induced by the double trace deformation. We will carry out the holographic computation for both the scalar field and the spinor field, and, in even dimension we will compute the same quantity also by conformal field theory methods.

{\bf Note added:} While our work was being completed, the paper \cite{Vecchi:2010dd} appeared, and we found some overlap between the results contained in there and our treatment of the boundary conditions as discussed in Section 2.
\section{Double trace deformations and boundary conditions}

Usually, boundary conditions on the fields in the gravity theory are imposed by hand. The pure conformal field theory corresponds to either Dirichlet boundary conditions or mixed boundary condition with a specific value of the coefficient. Displacing this coefficient by a certain amount is dual to introducing the double trace deformation, and then one has to carefully renormalize the coefficient and the boundary value of the field \cite{Hartman:2006dy}. All in all, the prescription is somewhat involved, especially the renormalization step.

It turns out that things are much more transparent if the fields are left free to vary, and the boundary conditions are imposed through the variational principle by adding appropriate boundary terms to the action. In fact, the form of these boundary terms has a direct connection to the form of the double trace deformation, and it is easily generalized to more complicated perturbations. Also, the renormalization is introduced in a more uniform way. Part of this approach was also introduced in \cite{Vecchi:2010dd}, but our treatment of the case $\Delta > d/2$ is substantially different.

%-----------------------------------------------------------------
\subsection{Scalar Field}
%-----------------------------------------------------------------

Let us now quickly review the properties of the two point function of the scalar operator $\mathcal O$ as computed with CFT methods. Then we will show how the same result arises from the gravity theory, by imposing appropriate boundary conditions  through the variational principle.

\subsubsection{The field theory side}

Let $\exval{\ \ }_f$ denote the expectation value in the perturbed CFT
\[
  \exval{\mathcal Q}_f = \exval{\mathcal Q\ e^{-\frac{f}{2} \int\!\! \dd^d x\, \mathcal O^2 (x)}}_{\scriptscriptstyle\mathrm{CFT}}\ ,
\]
and let $\Delta$ be the conformal dimension of the operator $\mathcal O$ . By introducing a Hubbard-Stratonovich auxiliary field and exploiting the large $N$ limit, one can obtain the two point function\footnote{The operator $\mathcal O$ is normalized so that $\exval{\mathcal O (x) \mathcal O(y)}_{\scriptscriptstyle\mathrm{CFT}}= \frac{1}{A(\Delta)|x - y|^{2 \Delta}}$ where $A(\Delta) = (4\pi)^{d/2} 2^{-2\Delta} \Gamma(d/2-\Delta)/\Gamma(\Delta)$}\cite{Gubser:2002vv}
\[
  \exval{\mathcal O(k)\mathcal O^\dag(q)}_f = (2 \pi)^d \delta(k - q) \frac{1}{f + k^{d-2\Delta} }\ .
\]

If $\Delta < d/2$ this form of the correlation function has a nice interpretation from the renormalization group point of view. It displays how the relevant operator $\mathcal O^2$ added to the conformal action starts a renormalization group flow away from the conformal point, which ends on another conformal point, where the operator has dimension $d-\Delta>d/2$. In fact, at high energy, we have
\[
  \exval{\mathcal O(k)\mathcal O^\dag(q)}_f \sim (2 \pi)^d \delta(k - q) k^{2\Delta - d}\ ,
\]
as one would expect from an operator of dimension $\Delta$, whereas, at low energy,
\[
  \exval{\mathcal O(k)\mathcal O^\dag(q)}_f \sim (2 \pi)^d \delta(k - q) \sqr{\frac{1}{f} -\frac{k^{d-2\Delta}}{f^2 } + \ldots}\ ,
\]
that is, in position space,
\[
  \exval{\mathcal O(x)\mathcal O(y)}_f \sim \frac{1}{f^2A(d-\Delta)}\frac{1}{ |x - y|^{2 (d-\Delta)}}\ ,
\]
as is appropriate for an operator of dimension $d - \Delta$.

This derivation of the correlation function still works when $\Delta > d/2$,  but this is an artifact of the large $N$ approximation. In fact, in this case, the perturbation is irrelevant, and it means that the CFT is at the end point of some RG flow. However, the RG flow is irreversible, there are many high energy theories that can flow to the same IR effective field theory (CFT + irrelevant operators). To be able to follow backwards the flow additional information is needed that is not obtainable from the IR CFT, because the irrelevant operator will couple to other unspecified operators, turning on a beta function for them. The large $N$ approximation discards the coupling of $\mathcal O$ to other operators, and gives the illusion of being able to follow backwards the RG flow.

%-----------------------------------------------------------------
\subsubsection{The gravity side}
%-----------------------------------------------------------------

Now let us review how the gravity theory can yield the same physics. The correlation function can be obtained from the generating functional
\[
  W[J,f] = \log \exval{e^{\int \dd^d x \sqr{ - \frac{f}{2}\mathcal O^2(x) + J(x) \mathcal O(x)}}}_{\scriptscriptstyle \mathrm{CFT}}\ .
\] 

Both the perturbation $f \mathcal O^2$ and the source term $J\mathcal O$ can be considered perturbations of the conformal fixed point, so let us at first focus to the case in which they are zero, and the theory is purely conformal. Then the dual geometry will be pure AdS space, for which we choose Poincar\'e coordinates\footnote{Here we set the AdS radius $L=1$.}
\[
  g = \frac{\dd r^2 + \sum_{i=1}^{d} \dd x_i^2}{r^2}\ .
\]

The gravity theory must contain a scalar field $\phi$, dual to the scalar operator $\mathcal O$, and an appropriate action for the scalar field is\footnote{Since we are interested in the two-point function, we may limit ourselves to the study of a quadratic action.}
\begin{align*}
  S_0[\phi] = & \int_\eps^\infty \dd r \int \dd^d x\ \sqrt{g} \sqr{\frac12 g^{\mu \nu} \partial_\mu \phi \partial_\nu \phi + \frac{1}{2}m^2\phi^2} +\\
  & +\rnd{\frac{d}{2}-\nu} \int \dd^d x\ \eps^{-d} \frac{1}{2} \phi^2(\eps, x)\ ,
\end{align*}
where
\[
  \nu = \sqrt{m^2 + \frac{d^2}{4}}\ .
\]

As usual, since the metric is singular at $r = 0$, it is necessary to cut off spacetime at $r = \eps$. Eventually we will take the limit $\eps \to 0$, with appropriate renormalization of some quantities. The boundary term, often called holographic renormalization term, is necessary to have conformal invariance in the dual theory\footnote{This is sufficient when $\nu \in [0,1]$. To extend to a wider range of masses additional holographic renormalization counterterms are needed.}. In fact, when it is not present, the on-shell action does not have a good $\eps \to 0$ limit, and, at the quantum level, the fluctuations of the field backreact on the metric, causing a deviation from AdS geometry\footnote{More on this in the section about the $c$-conjecture.}.

Now we can introduce the perturbations $f \mathcal O^2$ and $J\mathcal O$. This breaks conformal invariance, and consequently the geometry changes. However, since the dimension of $\mathcal O$ will turn out to be
\[
  \Delta = \Delta_- \equiv \frac{d}{2} -\nu  \leq \frac{d}{2}\ ,
\]
both perturbations are relevant, they become negligible in the u.v. limit, and we can still assume asymptotic AdS geometry. The perturbations are introduced as additional boundary terms in the action, so that the generating functional is

\[
  W_-[J,f] = \log \int \pathdd{\phi} e^{-S_-[\phi,J,f]}\ ,
\]
where
\[
  S_-[\phi,J,f] = S_0[\phi] + \int \dd^d x \sqr{ \rnd{\lambda \eps^{-\Delta_-}}^2 \frac{f}{2} \phi^2(\eps,x) - \lambda \eps^{-\Delta_-} J(x) \phi(\eps, x)}\ .
\]
and where the minus sign in the subscript reminds that this functional yields correlators of an operator with dimension $\Delta_-$. Later we will discuss how to obtain a conformal dimension greater than $d/2$.

The factors of $\eps ^{-\Delta_-}$ are renormalization factors, needed to have a good $\eps \to 0$ limit, and the factor $\lambda$ is a finite renormalization factor needed to exactly reproduce the field theory result. It is interesting to note that the same factor renormalizes $\phi$ in both terms. Apart from these renormalization issues, the way in which the perturbations of the conformal field theory are introduced is very transparent, and it is easily generalized to more complicated perturbations.

We use the saddle point approximation in evaluating the path integral, which amounts to disregarding subleading corrections in the large $N$ limit. We have
\[
  W_-[J,f] = - \min_{\phi}\ S_-[\phi,J,f]\ .
\]

We want to extremize $S_-$. Let us first look at the variation of $S_0$. We have
\begin{align*}
  S_0[\phi + \delta \phi] - S_0[\phi] = & \int \dd r \dd^d x\ \sqrt{g} \sqr{-g^{\mu \nu} D_\mu D_\nu \phi +m^2 \phi} \delta \phi\ -\\
  & -\int \dd^d x\ \eps^{-d} \pi(\eps, x) \delta \phi(\eps, x)\ ,
\end{align*}
where
\[
  \pi(r,x) = \eps^d \frac{\delta S_0}{\delta[\partial_r \phi(r,x)]} = \rnd{r \partial_r  - \frac{d}{2} + \nu} \phi(r, x)\ .
\]

The variation of the other terms is easy to compute and, putting all terms together we have
\begin{align*}
  &S_-[\phi+ \delta \phi, J, f] - S_-[\phi, J, f] =  \int \dd r\dd^d x\ \sqrt{g} \sqr{\mathrm{e.o.m.}} \delta \phi + \\
  &+ \int \dd^d x\eps^{-d}\sqr{- \pi(\eps,x) + \lambda^2 \eps^{2\nu} f \phi(\eps,x)  - \lambda \eps^{\frac{d}{2} +\nu}  J(x)}\delta \phi(\eps,x)\ .
\end{align*}

Then the field configuration $\phi_J$ that extremizes the action is the one that satisfies
\[
  \left\{
  \begin{array}{l}
    \displaystyle \vspace{0.2cm}\rnd{-g^{\mu \nu} D_\mu D_\nu +m^2} \phi_J(r, x) = 0 \\
    \displaystyle - \pi_J(\eps,x) +  \lambda^2 \eps^{2\nu} f \phi_J(\eps,x)  = \lambda \eps^{\frac{d}{2} +\nu}  J(x)\ ,
  \end{array}
  \right.
\]
so that
\[
  W_-[J,f] = -S_-[\phi_J, J,f]\ .
\]

Now we could compute explicitly the on shell action $S_-[\phi_J, J,f]$, by substituting the solution of the equations of motion. However, since we are mainly interested in the correlation function
\[
  \exval{\mathcal O(x) \mathcal O(y)}_f = \left. \frac{\delta^2 W_-}{\delta J(x) \delta J(y)}\right|_{J=0}\ ,
\]
we will vary the sources first, and substitute later. Let us look at the variation of $W_-$ when $J$ changes:
\begin{align*}
  W_-[J+\delta J,f]-W_-[J,f] & = - S_-[\phi_{J+\delta J}, J+ \delta J,f] + S_-[\phi_{J}, J,f]  \\
  & = \int \dd^d x \lambda \eps^{-\frac{d}{2} + \nu} \phi_J(\eps,x) \delta J(x)\ ,
\end{align*}
where we have used the fact that the action is at a stationary point with respect to variations of $\phi$. Then we have
\[
  \exval{\mathcal O(x) \mathcal O(y)}_f = \lambda \eps^{-\frac{d}{2} + \nu} \left. \frac{\delta \phi_J(\eps,x)}{\delta J(y)}\right|_{J=0}\ .
\]

To proceed further we need the small $r$ behavior of the solution $\phi_J$, so let us look at the equations of motion near the boundary
\[
  \sqr{-r^2\partial_r^2 +(d-1)r \partial_r - r^2\partial^2 + m^2}\phi(r,x) = 0\ .
\]

We can expand in plane waves in the transverse directions and find that the solution has the following small $r$ behavior
\[
  \phi_J(r,k) \sim A_J(k) \sqr{(kr)^{\frac{d}{2} -\nu} - a\ (kr)^{\frac{d}{2} +\nu}}
  \quad \mathrm{for} \quad
  r \to 0\ ,
\]
where the constant $a$ is fixed by the requirement that the solution be regular in the interior of space. For example, in the case of pure AdS geometry, we have\footnote{It is important that $a>0$ for the Green's function to be positive. In the case of pure AdS, $a$ is positive for $\nu \in [0,1]$, which is the range we are considering. It would be interesting to investigate more in general the conditions under which the Green's function is positive.}
\[
  a = - 2^{-2\nu} \frac{\Gamma(-\nu)}{\Gamma(\nu)} \ .
\]

Substituting the expansion in the definition of $\pi$ we have
\[
  \pi_J(r,k) \sim A_J(k) \sqr{- 2a\nu (kr)^{\frac{d}{2} +\nu}}
  \quad \mathrm{for} \quad
  r \to 0\ .
\]

The solution that satisfies the boundary condition is then easily obtained
\[
  \phi_J(r,k) = \lambda J(k) k^{-\frac{d}{2} + \nu} \frac{(kr)^{\frac{d}{2} -\nu} - a\ (kr)^{\frac{d}{2} +\nu}}{\lambda^2 f \sqr{1 - a(k\eps)^{2\nu}} + 2a\nu k^{2 \nu}}\ ,
\]
and we have
\begin{align*}
  \exval{\mathcal O(k) \mathcal O^\dag(q)}_f & = (2\pi)^d\lambda \eps^{-\frac{d}{2} + \nu} \left. \frac{\delta \phi_J(\eps,k)}{\delta J(q)}\right|_{J=0} \\
  & = (2\pi)^d \delta(k - q) \frac{\lambda^2 \sqr{1 - a (k \eps)^{2 \nu}}}{\lambda^2 f \sqr{1- a(k\eps)^{2\nu}} + 2a\nu k^{2 \nu}}\ .
\end{align*}

Setting $\lambda = \sqrt{2 a \nu}$ and taking the $\eps \to 0$ limit, we recover the field theory result with $\Delta = \Delta_-$
\[
  \exval{\mathcal O(k) \mathcal O^\dag(q)}_f= (2 \pi)^d \delta(k - q) \frac{1}{f + k^{d-2\Delta_-}}\ .
\]

How can the previous approach extend to operators of conformal dimension greater than $d/2$? First of all we remark that this is meaningful only when $f = 0$, otherwise we would be following backwards in the RG an irrelevant perturbation\footnote{As with the field theory computation, it is formally possible to make the perturbation $f$ work even when $\Delta>d/2$, by adding a $\pi^2$ term to the action, but we stress that this is an artifact of the large $N$ limit.}. Then we notice that the most naive approach would be to just send $\nu \to -\nu$. However it is easy to see that in this case the terms that were suppressed in the $\eps \to 0$ limit would become leading, yielding a trivial correlation function. What we have to do is to keep $\nu>0$ and add a new boundary term to the action, so that the generating functional becomes
\[
  W_+[J] = - \log \int \pathdd{\phi} e^{-S_+[\phi,J]}\simeq \min_{\phi}\ S_+[\phi,J]\ ,
\]
where
\begin{align*}
  S_+[\phi,J,f] =& S_0[\phi] + \int \dd^d x\sqr{ \eps^{-d} \pi(\eps,x )\phi(\eps,x) + \lambda \eps^{-\Delta_+} J(x) \pi(\eps, x)} \ .
\end{align*}

Varying this action we have
\begin{align*}
  &S_+[\phi+ \delta \phi, J] - S_+[\phi, J] =  \int \dd r\dd^d x\ \sqrt{g} \sqr{\mathrm{e.o.m.}} \delta \phi + \\
  &+ \int \dd^d x\eps^{-d}\sqr{\phi(\eps,x) + \lambda \eps^{\frac{d}{2} -\nu}  J(x)}\delta \pi(\eps,x)\ .
\end{align*}

The field configuration that extremizes the action is the one that satisfies
\[
  \left\{
  \begin{array}{l}
    \displaystyle \vspace{0.2cm}\sqr{-g^{\mu \nu} D_\mu D_\nu +m^2} \phi_J(r, x) = 0 \\
    \displaystyle  \phi_J(\eps,x) = - \lambda \eps^{\frac{d}{2} -\nu}  J(x)\ .
  \end{array}
  \right.
\]

Using the same approach as before we have
\[
\exval{\mathcal O(x) \mathcal O(y)}_f =  \lambda \eps^{-\frac{d}{2} - \nu} \left. \frac{\delta \pi_J(\eps,x)}{\delta J(y)}\right|_{J=0}\ ,
\]
and
\begin{align*}
  \exval{\mathcal O(k) \mathcal O^\dag(q)}_f = (2\pi)^d \delta(k - q) \frac{\lambda^2 2 a \nu\ k^{2\nu}}{1- a(k \eps)^{2\nu}} \ .
\end{align*}

Taking the $\eps \to 0$ limit and setting $\lambda = (2 a \nu)^{-1/2}$ we recover the field theory result with  $\Delta = \frac{d}{2} + \nu$ and $f = 0$.

%=================================================================
\subsection{Spinor field}
%=================================================================

We will now show how the results of the previous section extend to the case of a spinor field.

%-----------------------------------------------------------------
\subsubsection{The field theory side}
%-----------------------------------------------------------------

Consider a conformal field theory containing a spinor field $\chi$ of dimension $\Delta$. The partition function of the perturbed theory is
\[
  Z[\eta,\bar\eta;f] = Z_0 \exval{e^{-f \bar \chi \chi + \bar \eta \chi + \bar \chi \eta}}_{\scriptscriptstyle{\mathrm{CFT}}}\ ,
\]
where $\exval{\quad}_{\scriptscriptstyle{\mathrm{CFT}}}$ denotes the expectation value in the unperturbed field theory, and where we used the short hand notation\footnote{$h$ is the fixed background metric on which the field theory lives.}
\[
  \bar \chi \eta = \int \dd^d x \sqrt{h(x)}\ \bar \chi(x) \eta(x)\ .
\]

We want to compute this partition function exploiting the large $N$ limit, by the same trick used in \cite{Gubser:2002vv} for the scalar field. We introduce a couple of auxiliary, Grassmann-valued fields $\sigma$, $\bar \sigma$, and we have
\[
  Z[\eta,\bar\eta;f] = \tilde Z_0 \int \pathdd{\sigma}\pathdd{\bar\sigma} \exval{e^{\bar \sigma \sigma -f \bar \chi \chi + \bar \eta \chi + \bar \chi \eta}}_{\scriptscriptstyle{\mathrm{CFT}}}\ .
\]

The partition function is rescaled by an overall factor that does not depend on $f$ or $\eta$. Now we make the shift
\[
  \sigma \to \sigma + \sqrt{f} \chi\ ,\quad
  \bar \sigma \to \bar \sigma + \sqrt{f} \bar\chi\ ,\quad
\]
and we have
\[
  Z[\eta,\bar\eta;f] = \tilde Z_0 \int \pathdd{\sigma}\pathdd{\bar\sigma} \exval{e^{\bar \sigma \sigma + (\bar \eta+\sqrt{f} \bar \sigma) \chi + \bar \chi (\eta+\sqrt{f}\sigma)}}_{\scriptscriptstyle{\mathrm{CFT}}}\ .
\]

Now we use the fact that, for large $N$,
\[
  \exval{e^{\bar \eta\chi + \bar \chi \eta}} \simeq e^{\bar \eta G \eta}\ ,
\]
where $G$ denotes the convolution with $\exval{\chi \bar \chi}_{\scriptscriptstyle\mathrm{CFT}}$
\[
  (G \sigma)(x) = \int \dd^d y\ \exval{\chi(x) \bar \chi(y)}_{\scriptscriptstyle\mathrm{CFT}}\sigma(y)\ .
\]

Then we have
\[
  Z[\eta,\bar\eta;f] \simeq \tilde Z_0 \int \pathdd{\sigma}\pathdd{\bar\sigma} e^{\bar \sigma \sigma + (\bar \eta+\sqrt{f} \bar \sigma) G (\eta+\sqrt{f}\sigma)}\ .
\]

Now we make the shift
\[
  \sigma \to \sigma - \frac{\sqrt{f} G}{1+ f G} \eta\ ,\quad
  \bar \sigma \to \bar \sigma - \frac{\sqrt{f} G}{1+ f G} \bar \eta\ ,
\]
and we get
\begin{align}\label{eq::CFT_partition function}
  Z[\eta,\bar\eta;f] & = \tilde Z_0 \int \pathdd{\sigma}\pathdd{\bar\sigma} e^{\bar \sigma(1+fG) \sigma + \bar \eta\frac{G}{1+fG}\eta}\nonumber\\
  & = Z_0 \det(1+fG) e^{ \bar \eta\frac{G}{1+fG}\eta}\ .
\end{align}

From this, taking derivatives with respect to the sources, we have
\[
  \exval{\chi(x) \bar \chi(y)}_f = \frac{G}{1+fG}\delta(x-y) \ .
\]

On flat space we have
\[
  \exval{ \chi(x) \bar \chi(y)}_{\scriptscriptstyle\mathrm{CFT}} = \frac{1}{B(\Delta)}\frac{\gamma \cdot(x-y)}{|x-y|^{2\Delta + 1}}\ ,
\]
and we choose the normalization factor to be
\[
  B(\Delta) = (4 \pi)^{d/2} 2^{-\Delta} \frac{\Gamma(d/2-\Delta + 1/2)}{\Gamma(\Delta + 1/2)}\ .
\]

Then the operator $G$ is diagonal in the basis of momentum eigenstates
\[
  G(k,q) = (2\pi)^d \delta(k-q)  i \gamma \cdot \hat k\ k^{2\Delta-d}\ ,
\]
where $\hat k = k /|k|$, and we have
\[
  \exval{\chi(k) \bar \chi(q)}_f = (2\pi)^d \delta(k-q) \frac{1}{f - i \gamma \cdot \hat k\ k^{d-2\Delta}}\ .
\]

If $\Delta < d/2$ this correlation function can again be interpreted in terms of renormalization group flow from a conformal point in which the operator $\chi$ has dimension $\Delta$ to a conformal point in which the operator $\chi$ has dimension $d -\Delta$.

%-----------------------------------------------------------------
\subsubsection{The gravity side}
%-----------------------------------------------------------------

In the gravity theory, the operator $\chi$ is dual to a spinor field $\psi$. The action for the spinor field that is dual to a pure conformal field theory is\footnote{This action is valid for $m \in [0,1/2]$. To extend to a wider range of masses additional holographic renormalization counterterms are needed.}

\begin{align*}
  S_0[\psi, \bar \psi] = & \int_{\eps}^{\infty}\dd r \int \dd^d x\ \sqrt{g} \bar \psi \sqr{\frac12\rnd{\overrightarrow{\ovslash{D}} - \overleftarrow{\ovslash{ D}}} - m}\psi +\\
  & +  \int \dd^d x \eps^{-d}\frac12\left.\bar \psi \psi\right|_{r = \eps}\ .
\end{align*}

Also in this case, the holographic renormalization term is necessary to have a good $\eps \to 0$ limit \cite{Henneaux:1998ch} \nocite{Henningson:1998cd} \nocite{Mueck:1998iz}. This action will yield correlators of an operator with dimension
\[
  \Delta = \Delta_- \equiv d/2 -m < d/2\ ,
\]
later we will show how to obtain a conformal dimension larger than $d/2$.

When we introduce perturbations, i.e. we want to obtain the generating functional
\[
  W[\eta,\bar \eta, f] = \log Z[\eta,\bar \eta, f] = \log \exval{e^{\, \int\!\! \dd^d x \sqr{- f \bar \chi \chi + \bar \eta \chi + \bar \chi \eta}}}_{\scriptscriptstyle\mathrm{CFT}}\ ,
\]
we need to add some boundary terms to the action. We have
\[
  W_-[\eta,\bar \eta, f] = \log \int \pathdd{\psi}\pathdd{\bar \psi} e^{-S_-[\psi,\bar \psi ,\eta,\bar \eta, f]} \simeq - \min_\psi S_-[\psi, \bar \psi, \eta, \bar \eta, f]\ ,
\]
with\footnote{From now on we will not write explicitly the dependence on the conjugate fields.}
\[
  S_-[\psi,\eta,f] = S_0[\psi] + \int \dd^d x \sqr{(\lambda \eps^{-\Delta_-})^2 f \bar\psi \psi - \lambda \eps^{-\Delta_-} (\bar \eta \psi + \bar \psi \eta)}\ .
\]

We want to extremize this action, so let us first look at the variation of the action $S_0$
\begin{align*}
  S_0[\psi + \delta \psi]- S_0[\psi] & = \int \dd r \dd^d x \sqr{\delta \bar \psi (\overrightarrow{\ovslash{D}} - m) \psi + \bar \psi (-\overleftarrow{\ovslash{D}}-m)\delta \psi} + \\
  & + \int \dd^d x \eps^{-d} \sqr{\delta \bar \psi(\eps,x) \pi(\eps,x) + \bar \pi(\eps,x) \delta \psi(\eps,x)}\ ,
\end{align*}
where
\[
  \pi(r,x) = - \eps^{d} \frac{\delta S_0}{\delta[\partial_r \bar \psi(r,x)]} = P_+ \psi,
  \quad
  \bar \pi(r,x) = - \eps^{d} \frac{\delta S_0}{\delta[\partial_r \psi(r,x)]} = \bar \psi P_-,
\]
and
\[
  P_{\pm} = \frac{1 \pm \gamma^r}{2}\ .
\]

Varying the action $S_-$ we have
\begin{align*}
  S_-[\psi + \delta \psi,\eta]& - S_-[\psi,\eta]  = \mathrm{e. o. m.} \\
  & + \int \dd^d x \eps^{-d} \delta \bar \psi(\eps,x)\sqr{ \pi(\eps,x) + \lambda^2 \eps^{2m} f \psi(\eps,x) - \lambda \eps^{\frac{d}{2}+m}\eta(x)} \\
  & + \int \dd^d x \eps^{-d} \sqr{\bar \pi(\eps,x) + \lambda^2 \eps^{2m} f \bar \psi(\eps,x) - \lambda \eps^{\frac{d}{2}+m} \bar \eta(x)}\delta \psi(\eps,x)\ .
\end{align*}

The field configuration $\psi_\eta$, $\bar\psi_\eta$ that extremizes the action is the one that satisfies
\[
  \left\{
  \begin{array}{l}
    \displaystyle \vspace{0.2cm}\rnd{\overrightarrow{\ovslash{D}} - m} \psi_\eta(r, x) = 0 \\
    \displaystyle  \pi_\eta(\eps,x) + \lambda^2 \eps^{2m} f \psi_\eta(\eps,x)= \lambda \eps^{\frac{d}{2} + m}  \eta(x)\ ,
  \end{array}
  \right.
\]
and a similar set of equations for $\bar \psi$.

Varying the on-shell action $S_-$ with respect to $\eta$ and using the fact that the action is stationary for variations of $\psi$ and $\bar \psi$ we get
\[
  \exval{\chi(x) \bar \chi(y)}_f = \frac{\delta^2 W}{\delta \eta(y) \delta \bar\eta(x)} = \lambda \eps^{-\frac{d}{2}+m} \frac{\delta \psi_\eta(\eps,x)}{\delta \eta(y)}\ .
\]

To proceed further we need at least the asymptotic behavior for small $r$ of the solution to the equation of motion
\[
  \rnd{\ovslash{\overrightarrow{D}} - m} \psi = 0\ ,
\]
or, in coordinates
\begin{align*}
  &\sqr{\gamma^r \rnd{r\partial_r -\frac{d}{2}} + r \gamma \cdot \partial - m} \psi(r,x) = 0\ .
\end{align*}

To reduce clutter we will use the exact analytic solution in pure AdS space, instead of just the asymptotic behavior, but it is understood that the results do not rely on the assumption of having pure AdS background. 

It is easy to check by substitution that the most general solution to the equations of motion is
\begin{align*}
  \psi(r,k) = (kr)^{\frac{d+1}{2}}\sqr{K_{m-\frac{1}{2}}(kr) - K_{m + \frac{1}{2}}(kr)\ i \gamma \cdot \hat k} A(k)\ ,
\end{align*}
where $A(k)$ is a constant spinor that satisfies
\[
  \gamma^r A(k) = A(k)\ .
\]

Imposing the boundary conditions we have
\begin{align*}
  \psi_\eta(r,k) = & \lambda \eps^{\frac{d}{2}+m}\rnd{\frac{r}{\eps}}^{\frac{d+1}{2}}\times \\
  & \times \frac{K_{m-\frac{1}{2}}(kr) - K_{m + \frac{1}{2}}(kr)\ i \gamma \cdot \hat k}{K_{m - \frac{1}{2}}(k\eps) + \lambda^2 \eps^{2m} f \sqr{K_{m-\frac{1}{2}}(k\eps) - K_{m + \frac{1}{2}}(k\eps)\ i \gamma \cdot \hat k}}\ \eta(k) \ ,
\end{align*}
and hence
\begin{align*}
  \exval{\chi(k) \bar \chi(q)}_f = & (2\pi)^d \delta(k-q) \lambda^2 \eps^{2m} \times \\
  & \times \frac{K_{m-\frac{1}{2}}(k\eps) - K_{m + \frac{1}{2}}(k\eps)\ i \gamma \cdot \hat k}{K_{m - \frac{1}{2}}(k\eps) + \lambda^2 \eps^{2m} f \sqr{K_{m-\frac{1}{2}}(k\eps) - K_{m + \frac{1}{2}}(k\eps)\ i \gamma \cdot \hat k}}\ .
\end{align*}

To take the $\eps \to 0$ limit we need the asymptotic expansion of the modified Bessel function:
\[
  K_\nu(z) \sim a_\nu z^\nu + a_{-\nu} z^{-\nu}\ ,
  \quad
  a_\nu = 2^{-\nu-1}\Gamma(-\nu)\ .
\]

In the range $m \in [0,1/2]$ the leading contributions come from the terms $(k\eps)^{-m-\frac{1}{2}}$ and $(k\eps)^{m-\frac{1}{2}}$, so that, after some manipulations we have
\[
\exval{\chi(k) \bar \chi(q)}_f = (2\pi)^d \delta(k-q) \lambda^2 \sqr{\frac{a_{m-\frac{1}{2}}}{a_{-m-\frac{1}{2}}}k^{2m} i \gamma \cdot \hat k + \lambda^2 f}^{-1}\ ,
\]
and, upon setting
\[
  \lambda = \sqrt{\frac{a_{m-\frac{1}{2}}}{a_{-m-\frac{1}{2}}}}\ ,
\]
the field theory result is recovered with $\Delta = \Delta_-$.

To obtain a conformal dimension greater than $d/2$ we have to add a boundary term to the action, that is, we have to consider
\[
  W_+[\eta,\bar \eta, f] = - \log \int \pathdd{\psi}\pathdd{\bar \psi} e^{-S_+[\psi,\bar \psi ,\eta,\bar \eta, f]} \simeq \min_\psi S_+[\psi, \bar \psi, \eta, \bar \eta, f]\ ,
\]
where
\begin{align*}
  S_+[\psi,\eta] & = S_0[\psi] - \int\dd^d x \eps^{-d} \sqr{\bar \pi(\eps,x) \psi(\eps,x) + \bar \psi(\eps, x) \pi(\eps,x)}\\ 
  & + \int \dd^d x \lambda \eps^{-\Delta_+} \sqr{\bar \eta(x) \pi(\eps,x) + \bar \pi(\eps,x)\eta(x)}\ .
\end{align*}

Varying the action $S_+$ we have
\begin{align*}
  S_+[\psi + \delta \psi,\eta]& - S_+[\psi,\eta]  = \mathrm{e. o. m.} \\
  & + \int \dd^d x \eps^{-d} \delta \bar \pi(\eps,x)\sqr{- \psi(\eps,x) + \eps^{d/2-m} \eta(x)} \\
  & + \int \dd^d x \eps^{-d} \sqr{-\bar \psi(\eps,x) + \eps^{d/2-m} \bar \eta(x)}\delta \pi(\eps,x)\ .
\end{align*}

The field configuration $\psi_\eta$, $\bar\psi_\eta$ that extremizes the action is the one that satisfies
\[
  \left\{
  \begin{array}{l}
    \displaystyle \vspace{0.2cm}\rnd{\overrightarrow{\ovslash{D}} - m} \psi_\eta(r, x) = 0 \\
    \displaystyle  \psi_\eta(\eps,x) = \lambda \eps^{\frac{d}{2} - m}  \eta(x)\ .
  \end{array}
  \right.
\]

Varying the sources we have
\[
  \exval{\chi(x) \bar \chi(y)}_f = - \lambda \eps^{-\frac{d}{2}-m} \frac{\delta \pi_\eta(\eps,x)}{\delta \eta(y)}\ ,
\]
and the solution that satisfies the boundary conditions has
\begin{align*}
   \pi_\eta(r,k) =& \lambda \eps^{\frac{d}{2}-m}\rnd{\frac{r}{\eps}}^{\frac{d+1}{2}} \frac{K_{m-\frac{1}{2}}(kr)}{K_{m-\frac{1}{2}}(k\eps) - K_{m + \frac{1}{2}}(k\eps)\ i \gamma \cdot \hat k }\ \eta(k) \ ,
\end{align*}
so that
\begin{align*}
  \exval{\chi(k) \bar \chi(q)}_f = & -(2\pi)^d \delta(k-q) \lambda^2 \eps^{-2m} \frac{K_{m-\frac{1}{2}}(k\eps)}{K_{m-\frac{1}{2}}(k\eps) - K_{m + \frac{1}{2}}(k\eps)\ i \gamma \cdot \hat k }\ .
\end{align*}

Taking the $\eps \to 0$ limit we have
\[
  \exval{\chi(k) \bar \chi(q)}_f = (2\pi)^d \delta(k-q) \lambda^2 \sqr{\frac{a_{-m-\frac{1}{2}}}{a_{m-\frac{1}{2}}}k^{-2m} i \gamma \cdot \hat k }^{-1}\ ,
\]
and, upon setting
\[
  \lambda = \sqrt{\frac{a_{-m-\frac{1}{2}}}{a_{m-\frac{1}{2}}}}\ ,
\]
we recover the field theory result for $\Delta = d/2 + m$, $f = 0$.

%=================================================================
\section{The $c$-conjecture}
%=================================================================

We will now review some results on the holographic $c$-conjecture. According to Zamolodchikov's famous $c$-theorem \cite{Zamolodchikov:1986gt}, in a two dimensional field theory there exists a function of all the possible couplings that is monotonically decreasing along the trajectories of the renormalization group flow. Moreover, at the fixed points of the RG, where the theory is conformal, the value of this function coincides with the central charge of the CFT.

In another famous paper \cite{Cardy:1988cwa} Cardy conjectured that the same may be true for higher dimensional field theories, and proposed the following form for the function\footnote{The coefficient $a_d$ is a normalization factor. It is fixed by the requirement that, for a single free massless scalar field, $c = 1$, in any dimension. In particular, with our definition of the stress energy tensor, a free massless scalar field in two dimensions has $\exval{T_i^i} = - \frac{\mathcal R}{24 \pi}$ and hence $a_2 = 3$.}
\[
  c \equiv (-1)^{\frac{d}{2}} a_d \int_{S^d} \dd^d x\ \sqrt{h}\ h_{ij} \exval{\mathcal T^{ij}}\ ,
\]
where the field theory is supposed to live on an euclidean spherical background $S^d$, with round metric $h$ and where $\mathcal T$ is the stress energy tensor of the field theory.

In two dimensions, Cardy's definition reduces to Zamolodchikov's $c$-function, but his definition is interesting also in higher dimensions. In fact, in even-dimensional conformal field theories, the conformal symmetry is anomalous, and hence the $c$-function can take a non-trivial (non-zero) value at a fixed point of the RG\footnote{Often people still give to this value of $c$ the name of central charge, even when dimensionality is higher than two.}. Moreover Cardy gives evidence that the $c$-function may be monotonically decreasing along the RG trajectories\footnote{Recently a counterexample has been found \cite{Shapere:2008un} in 4 dimensions, in which this $c$-function increases. This probably means that the $c$-theorem in higher dimensions holds only under more stringent conditions than in two dimensions.} . For odd dimensional conformal field theories, instead, there is no conformal anomaly, and the $c$-function is just zero at every fixed point.

More recently, another possible definition of a $c$-function that works in any dimension and reduces to Zamolodchikov's definition in two dimensions has been proposed \cite{Myers:2010xs}. According to this proposal, one has to put the field theory on the spacetime $\reals \times S^{d-1}$, i.e. choose the boundary metric\footnote{In this paper $d$ always refers to the total number of dimensions of the boundary field theory.}
\[
  h = - \dd t^2 + R^2 \dd \Omega_{d-1}
\]
and compute the entanglement entropy in the ground state of half the sphere. A certain subleading universal contribution to this entanglement entropy, which we will discuss later, is conjectured to be decreasing along the trajectories of the RG-flow\footnote{At least under some suitable conditions, to exclude the counterexample \cite{Shapere:2008un}.}.

The renormalization group flow induced by double trace deformations is a case in which the $c$-conjecture and the various proposals of $c$-functions can be tested. In the case of perturbation by a scalar operator, it has been possible to compute the change in the value of Cardy's $c$-function between the two end points of the flow, by both holographic \cite{Gubser:2002zh} and CFT \cite{Gubser:2002vv} methods, showing that it is negative. 

In the following we will show how to compute the $c$-function from the holographic principle, review the results on the scalar field and extend them to the spinor field. We will show that the $c$-function is decreasing, in any dimension, for both the scalar and the spinor field. In addition, we will show that the same results can obtained, in even dimension, by CFT methods.

%-----------------------------------------------------------------
\subsection{Cardy's $c$-function from holography}
%-----------------------------------------------------------------

Let us briefly review how the holographic computation of Cardy's $c$-function can be accomplished \cite{Henningson:1998gx}\cite{Hartman:2006dy}. The partition function $Z$ of the field theory can be computed from the dual gravity theory, following the holographic principle. 

In a semiclassical approach in which the metric is treated classically and the matter content is given a full quantum treatment, we can write the gravity action as\footnote{Since we are in euclidean space, curvature must come with the negative sign, because the Wick rotation of the Minkowski action has to be consistent with what is conventionally done with the matter action \cite{Gibbons:1978ac}.}
\[
  S_{\mathrm {gr}}[g] = \frac{1}{16 \pi G} \int_\eps \dd r \int \dd^d x \sqrt{g} \rnd{-\mathcal R + 2 \Lambda + 16 \pi G V[g]}\ ,
\]
where $r$ is the holographic direction, the boundary being at $r = \eps$, and where $V$ is the effective potential\footnote{The effective potential $V$ is a divergent quantity, but its divergences are proportional to geometric quantities, and can be absorbed in the renormalization of the correspondent terms in the gravity action This usually brings about higher derivatives terms, but here we will stick to Einstein's gravity.}
\[
  \int_\eps \dd r \int \dd^d x \sqrt{g}\ V[g] = - \log \int \pathdd{\phi} e^{-S[\phi,g]}\ .
\]
where $\phi$ represents the matter content of the theory.

Then $- \log Z$ is given by the on-shell value of the gravity action, with the boundary condition that, at $r = \eps$, the transverse part of the metric coincides with the background metric $h$ of the dual field theory
\[
  \eps^2 g_{ij}(\eps, x) = h_{ij}(x)\ .
\]

In general, the matter content can act as a source for the gravitational field in an arbitrarily complicated way, but, for certain choices of the matter action, the AdS geometry turns out to be an extremum of $V[g]$. This is the case, in particular, for the actions discussed before, when both the perturbation and the source term are set to zero. In these particular cases, if $2\Lambda + 16 \pi G V < 0$, the geometry that extremizes the entire gravity action is the AdS geometry, and hence the dual field theory is a conformal field theory in its ground state.

Adding a double trace deformation to the dual field theory breaks conformal invariance and triggers a renormalization group flow. Correspondingly, in the gravity theory, adding a boundary term to the matter action yields an effective potential that is not extremized by AdS geometry. As a consequence, also the geometry that extremizes the entire gravity action deviates from pure AdS. If one could compute the new geometry, as a function of the renormalized parameter $f$, he could have a full geometric picture of the RG flow. In particular, he could obtain $c(f)$, and explicitly check that it is monotonously decreasing. However, this program is probably too ambitious. What is possible to do is to compare the end points of the RG flow, i.e. compute the change in quantities when the matter action changes from the $\Delta_-$ action to the $\Delta_+$ action.

In particular, we are interested in computing the change in the value of $c$ between the two conformal points, so we want the boundary metric to be the spherical metric
\[
  h = R^2 \dd \Omega_d^2\ .
\]

Then we have
\begin{align*}
  R \frac{\partial}{\partial R}(- \log Z) &= \int \dd^d x\ R\frac{\partial h_{ij}(x)}{\partial R} \frac{\delta}{\delta h_{ij}(x)}(- \log Z) \\
  &=\int \dd^d x\ 2\ h_{ij}(x)\ \frac{1}{2}\sqrt{h(x)}\exval{\mathcal T^{ij}(x)}\ ,
\end{align*}
that is
\[
  c = (-1)^{\frac{d}{2}} a_d R \frac{\partial}{\partial R}(- \log Z)\ .
\]

The metric that solves Einstein's equations with the given boundary conditions is the metric of AdS space in hyperbolic coordinates\footnote{In this section of the paper, we find convenient to write explicitly the AdS radius.},
\[
   g = \frac{L^2}{r^2}\dd r^2 + \frac{R^2}{r^2}\rnd{1 - \frac{L^2 r^2}{4 R^2}}^2 \dd \Omega_d^2\ ,
\]
where
\[
  L^2 = - d(d-1)\frac{1}{2\Lambda + 16 \pi G V}\ .
\]

Clearly the scalar curvature ${\mathcal R}$ of the AdS metric is constant, and we have
\[
  -{\mathcal R} + 2 \Lambda + 16 \pi G V = \frac{2d}{L^2}\ .
\]

The on-shell action then is
\begin{align*}
  -\log Z  & =  S_{\mathrm {gr}}[ g] = \frac{1}{16 \pi G}\frac{2d}{L^2} \int_\eps^{\frac{2R}{L}} \dd r \int \dd^d x \sqrt{g} \\
  & = \frac{1}{16 \pi G}\frac{2d}{L^2} \int_\eps^{\frac{2R}{L}} \dd r \int \dd \Omega_d\  \frac{L R^d}{r^{d+1}} \rnd{1-\frac{L^2 r^2}{4 R^2}}^d \\
  & = 2^{1-d} d\ \Sigma_d \frac{L^{d-1}}{16 \pi G} \int_{\frac{L\eps}{2R}}^{1} \dd r  r^{-d-1} \rnd{1-r^2}^d\ ,
\end{align*}
where $\Sigma_d$ is the surface of the $d$-sphere.

The integrand contains negative powers of $r$, and the integral is clearly not finite in the limit of $\eps \to 0$. The action needs to be regularized by adding some $\eps$-dependent boundary counterterms\footnote{ It can be shown that this can be done in a fully covariant way \cite{Henningson:1998gx}.}. The dependence on $R$ is only in the integration limit, in the form $\eps L/R$. After regularization and the $\eps \to 0$ limit, this dependence must drop, except for a possible term $\log R/L$, that could come from integrating the term proportional to  $1/r$. Such a term is present only for even $d$. Explicitly, in even $d$, the $R$ dependent term is
\[
  -\log Z \rightarrow  (-1)^{d/2}  2^{1-d} d\ \Sigma_d \binom{d}{\frac{d}{2}} \frac{L^{d-1}}{16 \pi G} \log R\ ,
\]
and hence
\[
  c = 2^{1-d} d\ \Sigma_d \binom{d}{\frac{d}{2}} a_d \frac{L^{d-1}}{16 \pi G}\ ,
\]
where $\Sigma_d$ is the surface of the $d$-sphere. In particular, in two dimension, this reproduces the well known result \cite{Brown:1986nw}
\[ 
  c = \frac{3 L}{2G}\ .
\]

When the effective potential changes by an amount 
\[
  \Delta V = V_+ - V_-\ ,
\]
the relative change in central charge is
\begin{align*}
  \frac{\Delta c}{c} & \simeq (d-1) \frac{\Delta L }{L} \simeq -\frac{d-1}{2}\frac{16 \pi G \Delta V }{2\Lambda + 16 \pi G V} \\
  & = \frac{1}{2d} 16 \pi G L^2 \Delta V = \frac{1}{N^2} \frac{L^{d+1}\Delta V}{2d}\ ,
\end{align*}
where we have discarded quantities of higher order in\footnote{This expression for $N$ is true in $\mathcal N = 4$ SYM theory. The coefficient may change in other theories with gravity duals.}
\[
  \frac{16 \pi G}{ L^{d-1}}= \frac{1}{N^2}\ .
\]

%-----------------------------------------------------------------
\subsection{Entanglement entropy $c$-function from holography}
%-----------------------------------------------------------------

According to the proposal of \cite{Myers:2010xs}, we choose the boundary metric to be\footnote{In this paper, $d$ will always refer to the total number of dimensions of the boundary field theory.}
\[
  h = - \dd t^2 + R^2 \dd \Omega_{d-1}\ ,
\]
and the metric that satisfies Einstein equations with negative cosmological constant and this boundary condition is
\[
   g = \frac{L^2}{r^2}\dd r^2 + \frac{1}{r^2}\crl{-\rnd{1 + \frac{L^2 r^2}{4 R^2}}^2\dd t^2 + {R^2}\rnd{1 - \frac{L^2 r^2}{4 R^2}}^2 \dd \Omega_{d-1}^2}\ .
\]

We want to compute the entanglement entropy of half the sphere. This is proportional to the area of the minimal surface in the bulk whose boundary is the equator of the sphere \cite{Ryu:2006ef}. The minimal surface is clearly the disk that cuts the AdS space in two. The metric induced on this surface is
\[
  k = \frac{L^2}{r^2}\dd r^2 + \frac{R^2}{r^2}\rnd{1 - \frac{L^2 r^2}{4 R^2}}^2 \dd \Omega_{d-2}^2\ ,
\]
and its area is
\[
  A = \int_\eps^{\frac{2R}{L}} \dd r \int \dd^{d-2} x \sqrt{k}\ .
\]

We are then instructed to discard all the contributions that are power divergent when $\eps \to 0$. When $d$ is odd the result is finite, when $d$ is even there is still a divergence proportional to $\log \eps$. The proposed $c$ function is the finite term in the first case, and the coefficient that multiplies $\log \eps$ in the second. It is apparent that the computation is identical, modulo an overall positive constant, to the one for Cardy's $c$-function, but this proposal also gives a meaning to the constant term that appears when $d$ is odd. Since the area $A$ is proportional to $L^{d-1}$, exactly as Cardy's $c$ function, we find that the $c$-function constructed from the entanglement entropy also changes as
\[
  \frac{\Delta c}{c} =\frac{1}{N^2} \frac{L^{d+1}\Delta V}{2d}\ ,
\]
for any $d$.

%-----------------------------------------------------------------
\subsection{Change in central charge from holography - scalar field}
%-----------------------------------------------------------------

Let us now review how the change in the effective potential can be computed for the case of a scalar field \cite{Gubser:2002zh}. According to our definition we have\footnote{Here we use a quadratic action, i.e. we compute the effective potential at one loop. Moreover, for brevity we drop the boundary terms in the action. However, they still control the boundary conditions on the fields, and this will become important later in the derivation.}
\[
  \int \dd r \dd^d x \sqrt{g}\ V = - \log \int \pathdd{\phi} e^{-\int \dd r \dd^d x \sqrt{g}\  \frac{1}{2}\sqr{(\partial \phi)^2 + \frac{m^2}{L^2} \phi^2}}\ .
\]

Since we are computing a density, we are free to use any coordinates we prefer to parametrize the space, even if they don't cover the whole space. Using Poincar\'e coordinates
\[
  g = \frac{L^2}{r^2}\rnd{\dd r^2 + \sum_{i=1}^{d} \dd x_i^2}
\]
is the most convenient choice. 

We are interested in computing the difference
\[
  \Delta V = V_+ - V_-\ .
\]
Computing $V$ directly is difficult, it is much easier to compute the derivative of $V$ with respect to $m^2$. At the Breitenlohner-Freedman bound $m^2 = -d^2/4$ the two asymptotic behaviors $\phi \sim r^{\Delta_{\pm}}$ of the fields become identical, and it can be argued \cite{Gubser:2002zh} that also the respective potentials become equal. Then we have
\[
  \Delta V(m^2) = \int_{-d^2/4}^{m^2} d(\mu^2) \Delta V'(\mu^2)\ ,
\]
where
\[
  \Delta V'(\mu^2) \equiv \left.\frac{\dd V_+}{\dd(m^2)}\right|_{m = \mu} - \left.\frac{\dd V_-}{\dd(m^2)}\right|_{m = \mu}\ .
\]

Taking the derivative with respect to $m^2$ of both sides of the equation defining $V$ we have
\[
  \int \dd r \dd^d x \sqrt{g}\ \totder{V_\pm}{(m^2)} = \frac{1}{2L^2} \int \dd r \dd^d x \sqrt{g} \exval{\phi(r,x)^2}\ ,
\]
that is\footnote{Because of the symmetry of AdS space, the effective potential does not depend on $r$, $x$.}
\[
  \totder{V_\pm}{(m^2)} = \frac{1}{2L^2} G_\pm(r,x;r,x)\ ,
\]
where $G$ is the propagator of the theory. Its computation is reported in the appendix, it has the form
\[
  G_\pm(r,x;s,y) = \frac{r^{\frac{d}{2}}s^{\frac{d}{2}}}{L^{d-1}}\int \frac{d^d k}{(2\pi)^d} e^{i k\cdot(x-y)} \int \dd \omega\ \omega\ \frac{J_{\pm\nu}(\omega r)J_{\pm\nu}(\omega s)}{\omega^2 + k^2}\ .
\]

Then we have
\[
  \totder{V_+}{(m^2)} = \frac{1}{2 L^{d+1}}\int \frac{d^d k}{(2\pi)^d} \frac{1}{1+k^2}\int \dd \omega\ \omega^{d-1} J_\nu(\omega )^2\ ,
\]
where we have rescaled the integration variables appropriately. Both integrals can be done exactly in dimensional regularization, and we have
\[
  \totder{V_+}{(m^2)} = \frac{1}{2L^{d+1}}(4\pi)^{-\frac{d+1}{2}} \Gamma\rnd{\frac{1-d}{2}} \frac{\Gamma\rnd{\frac{d}{2}+\nu}}{\Gamma\rnd{1-\frac{d}{2}+\nu}}\ .
\]

Now we subtract the expression with $\nu \to - \nu$ and set $d = \mathrm{integer} - 2 \eps$. The divergent contributions cancel and we obtain a finite expression for $\Delta V'(m^2)$. After a few manipulations we get
\[
  \Delta V'(m^2) = - \frac{1}{2(d-1)!\Sigma_d L^{d+1}}\ \nu \prod_{p=1}^{\frac{d-2}{2}}(p-\nu)(p+\nu)
\]
for even $d$ and
\[
  \Delta V'(m^2) = - \frac{1}{2(d-1)!\Sigma_d L^{d+1}}\ \tan (\pi \nu) \prod_{p=1}^{\frac{d-1}{2}}\rnd{p-\frac{1}{2} - \nu}\rnd{p - \frac{1}{2} +\nu}
\]
for odd $d$.

Now we should integrate in $m^2$ from the Breitenlohner-Freedman bound $\nu = 0$. However, here we are mainly interested in showing that $\Delta V < 0$, and this is already evident. In fact, at $\nu =0$, $\Delta V' <0$, and the first zero of $\Delta V'$ is at $\nu = 1$. Therefore, at least for $0 < \nu \leq 1$, that is
\[
  \frac{d}{2} < \Delta_+ \leq \frac{d+2}{2}\ ,
  \quad\quad
  \frac{d-2}{2} \leq \Delta_- < \frac{d}{2}\ ,
\]
we have $\Delta V < 0$.

%-----------------------------------------------------------------
\subsection{Change in central charge from holography - spinor field}
%-----------------------------------------------------------------

Now we will carry out a similar computation for the spinor field. In this case the relation between $V_-$ and $V_+$ is quite explicit. When the sources are turned off, we have
\[
  S_{+}[\bar\psi,\psi, m] = S_{-}[\bar\psi,\psi, m] - \int \dd^d x \eps^{-d}\sqr{\bar \pi(\eps,x) \psi(\eps,x) +\bar \psi(\eps,x) \pi(\eps,x) }\ ,
\]
and it is easy to check that
\[
  S_{+}[\bar\psi,\psi, m] = S_{-}[\psi^\dag, \bar \psi^\dag, -m]\ .
\]

Since the fields are integrated over, we have
\[
  V_+(m) = V_-(-m)\ ,
\]
and we are interested in the difference between the effective potentials. We have:
\[
  \Delta V(m) = V_+(m) - V_-(m) = \int_0^m \dd \mu\ \Delta V'(\mu)\ ,
\]
where
\begin{align*}
  \Delta V'(\mu) & = \totder{}{m}\sqr{V_+(m)- V_+(m)}_{m=\mu} \\
  & = \left. \totder{V_+}{m}\right|_{m=\mu} + \left.\totder{V_+}{m}\right|_{m=-\mu}\ .
\end{align*}

According to our definition, we have\footnote{Also here, for shortness, we dropped the boundary terms in the action. They will play an important role later when we will discuss the boundary conditions, but no role in the present derivation.}
\[
  \int \dd r  \dd^d x \sqrt{g}\ V_+[g] = - \log \int \sqr{\mathcal D \bar \psi \mathcal D \psi} e^{-\int \dd r \dd^d x \sqrt{g}\bar \psi\rnd{D\hspace{-1.9mm}\slash - \frac{m}{L}}\bar \psi}\ .
\]

Taking the derivative with respect to $m$ of both sides we have
\begin{align*}
  \totder{V_+}{m} & = - \frac{1}{L}\exval {\bar \psi(r,x) \psi(r,x)} =\frac{1}{L} \Tr\sqr{\exval{\psi(r,x) \bar \psi(r,x)}} \\
  & = \frac{1}{L}\Tr\sqr{G(r,x;r,x)}\ ,
\end{align*}
where $G$ is the propagator of the field theory. The computation of the propagator is reported in the appendix, it has the form
\begin{align*}
  & G(r,x;s,y) = - \frac{r^{\frac{d+1}{2}}s^{\frac{d+1}{2}}}{L^{d}}\int \frac{\dd^d k}{(2\pi)^d} e^{ik \cdot(x-y)} \int_0^\infty \dd \omega\ \omega\ \times\\
  &\times \sqr{ J_{m-\frac12}(\omega r) P_+ + J_{m+\frac12}(\omega r) P_-} \frac{\omega + i\gamma \cdot k }{k^2 + \omega^2}\sqr{P_+ J_{m+\frac12}(\omega s) + P_- J_{m-\frac12}(\omega s)}\ .
\end{align*}

Therefore, rescaling the integration variables appropriately we have
\[
  \totder{V_+}{m} = - \frac{\dim \gamma}{L^{d+1}} \int \frac{\dd^d k}{(2\pi)^d} \frac{1}{1+k^2} \int_{0}^\infty \dd \omega\ \omega^d\ J_{m-\frac12}(\omega) J_{m+\frac12}(\omega)\ ,
\]
where $\dim \gamma$ is the dimension of the representation of the Clifford algebra.

Both integrals can be done exactly, to get
\[
  \totder{V_+}{m} = - \frac{\dim \gamma}{L^{d+1}}(4\pi)^{-\frac{d+1}{2}}\Gamma\rnd{\frac{1-d}{2}}\frac{\Gamma\rnd{\frac{1+d}{2} + m}}{\Gamma\rnd{\frac{1-d}{2}+m}}\ .
\]

Now we add the expression with $m \to - m$, set $d = \mathrm{integer} - 2 \eps$, the divergent contributions cancel and we obtain a finite expression for $\Delta V'(m)$. After a few manipulations we get
\[
  \Delta V'(m) = - \frac{\dim \gamma}{(d-1)!\Sigma_d L^{d+1}}\ \prod_{p=1}^{\frac{d}{2}}\rnd{p-\frac{1}{2} - m}\rnd{p - \frac{1}{2} +m}
\]
for even $d$ and
\[
  \Delta V'(m) = - \frac{\dim \gamma}{(d-1)!\Sigma_d L^{d+1}}\ \cot (\pi m) m \prod_{p=1}^{\frac{d-1}{2}}\rnd{p - m}\rnd{p + m}
\]
for odd $d$.

From this expression it is already manifest that $\Delta V$ is negative. In fact, at $m =0$, $\Delta V' <0$, and the first zero of $\Delta V'$ is at $m = 1/2$. Therefore, at least for $0 < m \leq 1/2$, that is
\[
  \frac{d}{2} < \Delta_+ \leq \frac{d+1}{2}\ ,
  \quad\quad
  \frac{d-1}{2} \leq \Delta_- < \frac{d}{2}\ ,
\]
we have $\Delta V < 0$. 

For comparison with the CFT computation of the change in central charge, table \ref{tab::AdS_computation} reports the value of
\[
    \frac{\Delta c}{c} =\frac{1}{N^2} \frac{L^{d+1}\Delta V}{2d}
\]
for the first few even dimensions.

\begin{table}[htb]
\begin{center}
\begin{tabular}{|l|l|l|}
  \hline
  \rule[-2mm]{0pt}{7mm} d 
      &  $ N^2\Delta c/c$ \\
  \hline 
  \rule[-5mm]{0pt}{13mm} 2 
      & $\displaystyle -\frac{1}{16\pi}\rnd{m - \frac{4}{3}m^3}$ \\
  \rule[-5mm]{0pt}{13mm} 4 
      & $\displaystyle -\frac{3}{128\pi^2}\rnd{m - \frac{40}{27}m^3+ \frac{16}{45}m^5}$ \\
  \rule[-5mm]{0pt}{13mm} 6 
      & $\displaystyle -\frac{5}{256\pi^3}\rnd{m - \frac{1036}{675}m^3+ \frac{112 }{225}m^5 - \frac{64}{1575}m^7}$\\
  \hline
\end{tabular}
\end{center}
\caption{Relative change in central charge as computed from the gravity theory.}\label{tab::AdS_computation}
\end{table}

%-----------------------------------------------------------------
\subsection{Change in central charge from CFT methods - spinor field}
%-----------------------------------------------------------------

Let us now see how the same results can be obtained directly from the conformal field theory (we follow closely \cite{Gubser:2002vv}). According to our large $N$ computation of the partition function we have
\[
  Z[f]  = Z_0 \det(1+fG)\ ,
\]
where
\[
  G(\theta,\theta') = \exval{\chi(\theta) \bar \chi(\theta')}_{\scriptscriptstyle\mathrm{CFT}}\ .
\]

Now, according to Cardy's proposal for the $c$-function:
\begin{align*}
  c(f) - c(0) & = (-1)^{\frac{d}{2}}R \frac{\partial}{\partial R} (- \log Z[f]) \\
  & =(-1)^{\frac{d}{2}}\sqr{- R \frac{\partial}{\partial R}\log \det(1+fG)}\ ,
\end{align*}
where the $R$-dependence comes from the operator $G$.

In order to compute the determinant we will diagonalize $G$ and obtain eigenvalues and degeneracies. The simplest conformal field theory with a spinor field is the free Dirac field, and we assume that $G$ has the same symmetry properties of the Green's function of the Dirac operator. Hence, since the space is compact, the spectrum is discrete. Since the operator is antihermitian, the eigenvalues are pure imaginary numbers, and we assume that they come in conjugate pairs $\pm i g_n$. With this information, and defining $M^{(d)}(n)$ to be the degeneracy of each eigenspace, we have
\[
  c(f) - c(0) = (-1)^{\frac{d}{2}} \sqr{-R \frac{\partial}{\partial R} \sum_n M^{(d)}(n)\log \sqr{1+(fg_n)^2}}\ .
\]

Computing the $c$-function for generic $f$ presents difficulties of the same degree as the corresponding holographic computation. We will do what is possible and consider the limit of this expression for $f \to \infty$. This amounts to compute the difference in the central charge between the end points of the $RG$ flow:
\[
  (-1)^{\frac{d}{2}}\Delta c = - \lim_{f \to \infty} R \frac{\partial}{\partial R} \sum_n M^{(d)}(n) \log (fg_n)^2\ .
\]

The term proportional to $\log f$ in this expression does not depend on $R$ or $\Delta$, and therefore does not survive\footnote{This can be checked explicitly later when a regulated sum is introduced.} the derivative with respect to $R$. Then we have
\[
  (-1)^{\frac{d}{2}}\Delta c =  - 2 R \frac{\partial}{\partial R} \sum_n M^{(d)}(n)\log g_n\ .
\]

To proceed further we need an explicit expression for the spectrum of $G$ and the degeneracies. Their computation is reported in the appendix, they have the form
\[
  g_n \propto R^{d-2\Delta}\frac{\Gamma\rnd{n+\Delta +\frac{1}{2}}}{\Gamma\rnd{n+d-\Delta +\frac{1}{2}}}\ ,
\]
\[
  M^{(d)}(n) = \dim\gamma \frac{(n+d-1)!}{n!(d-1)!}\ ,
\] 
where $\dim\gamma$ is the dimension of the representation of the Clifford algebra.

Since our starting point, the unperturbed CFT, must have $\Delta < d/2$, let us set $\Delta = \frac{d}{2} - m$. We have to compute
\[
  (-1)^{\frac{d}{2}}\Delta c =  -2 R \frac{\partial}{\partial R} \sum_{n=0}^{\infty} M^{(d)}(n) \log\sqr{R^{2m}\frac{\Gamma\rnd{n +\frac{d+1}{2}-m}}{\Gamma\rnd{n +\frac{d+1}{2}+m}}}\ .
\]

The sum is divergent, and needs to be regulated. In general, there will be two contributions to the logarithmic derivative $R \partial_R$ \cite{Gubser:2002vv}. One will come from the term $\log R$ explicitly present:
\[
  (-1)^{\frac{d}{2}}\Delta c_1 = -4m \sum_{n=0}^{\infty} M^{(d)}(n)\ ,
\]
the other from the term in the sum
\[
   (-1)^{\frac{d}{2}}\Delta c_2 = - 2 R \frac{\partial}{\partial R}\sum_{n=0}^{\infty} M^{(d)}(n) \log\frac{\Gamma\rnd{n +\frac{d+1}{2}-m}}{\Gamma\rnd{n +\frac{d+1}{2}+m}}
\]
that is logarithmically divergent. In fact, any regulator will refer to some fixed energy scale $\Lambda$, so that such divergent contribution to the sum would be proportional to $\log R\Lambda$, and give a finite contribution to the logarithmic derivative. We will use zeta function regularization and a couple of words must be spent on how to extract either contributions from this regulator. By means of an appropriate shift\footnote{It is crucial that the shift be the same for both sums.} of $n$ and using some relations between special functions, we will cast the sums in the form
\begin{align*}
  \Delta c_1 & = \sum_{k=0}^{\infty} \sum_{r=0}^{d-1} a_r k^r \\
  \Delta c_2 & = R \frac{\partial}{\partial R}\sum_{k=0}^{\infty} \sum_{r=0}^{\infty} b_r k^{d-r-1}\ .
\end{align*}

The first one will be evaluated with the zeta function,
\[
  \Delta c_1 = \sum_{r=0}^{d-1} a_r \zeta(-r)\ .
\]

In the second one, the derivative with respect to $R$ extracts the coefficient of $\log \Lambda R$. When using the zeta function regularization, this coefficient is the coefficient that multiplies $1/k$ in the sum, i.e. $b_{d}$. In the following we will keep the derivative $R \partial_R$ with this meaning of an operator extracting the appropriate coefficient from the series.

Let us now show how to carry out what outlined above. We have
\[
  \log\frac{\Gamma\rnd{n +\frac{d+1}{2}-m}}{\Gamma\rnd{n +\frac{d+1}{2}+m}} = - 2 \sum_{p=0}^{\infty} \frac{m^{2p+1}}{(2p+1)!} \psi^{(2p)}\rnd{n+\frac{d+1}{2}}\ ,
\]
where $\psi(z) = \Gamma'(z)/\Gamma(z)$ is the digamma function. We will use the following asymptotic expansion
\[
  \psi(z) = \log(z) + \sum_{s=0}^{\infty} \zeta(-s) z^{-s-1}\ ,
\]
and the duplication relation
\[
  \psi\rnd{z + \frac{1}{2}} = 2 \psi(2 z)-\psi(z) -2\log 2\ .
\]

For $d$ even we can set $n = k - d/2$, and we have
\begin{align*}
  (-)^{\frac{d}{2}}\Delta c_1 & =  -4m \sum_{k=0}^{\infty} M^{(d)}\rnd{k-\frac{d}{2}} \\
  (-)^{\frac{d}{2}}\Delta c_2 &=   4R \frac{\partial}{\partial R}\sum_{k=0}^{\infty} M^{(d)}\rnd{k-\frac{d}{2}} \sum_{p=0}^{\infty} \frac{m^{2p+1}}{(2p+1)!} \psi^{(2p)}\rnd{k+\frac{1}{2}}\ ,
\end{align*}
where the sum can start from $k = 0$ and not from $k = d/2$ because $M^{(d)}(k-d/2)$ is zero for all the additional terms. The summand in $c_1$ is already a polynomial in $k$. We can use the asymptotic expansion of $\psi$ to express also $\Delta c_2$ as a power series in $k$. Then we have to compute $\Delta c_1$ with the zeta function regularization and to extract the coefficient that multiplies $1/k$ in the series for $\Delta c_2$. This is best done with the aid of the computer. Table \ref{tab::CFT_computation} reports the results for the first few dimensions, in agreement with the results from the gravity theory reported in table \ref{tab::AdS_computation} up to an overall positive constant.

\begin{table}[htb]
\begin{center}
\begin{tabular}{|l|l|}
  \hline
  d & $\Delta c$ \\
  \hline
  \rule[-5mm]{0pt}{12mm} 2 &
      $\displaystyle -\rnd{m - \frac{4}{3}m^3}$ \\
  \rule[-5mm]{0pt}{12mm} 4 &
      $\displaystyle -\frac{3}{8} \rnd{m - \frac{40}{27}m^3 + \frac{16}{45} m^5}$ \\
  \rule[-5mm]{0pt}{12mm} 6 &
      $\displaystyle -\frac{5}{32} \rnd{m - \frac{1036}{675}m^3 + \frac{112}{15} m^5 -\frac{64}{1575} m^7 }$ \\
  \hline 
\end{tabular}
\caption{Change in central charge as computed from the conformal field theory.}\label{tab::CFT_computation}
\end{center}
\end{table}

In odd dimensions we set $n = k - (d+1)/2$, and we have
\begin{align*}
  (-)^{\frac{d}{2}}\Delta c_1 & = - 4m \sum_{k=0}^{\infty} M^{(d)}\rnd{k-\frac{d+1}{2}}\ ,\\
  (-)^{\frac{d}{2}}\Delta c_2 & =  4R \frac{\partial}{\partial R}\sum_{k=0}^{\infty} M^{(d)}\rnd{k-\frac{d+1}{2}} \sum_{p=0}^{\infty} \frac{m^{2p+1}}{(2p+1)!} \psi^{(2p)}\rnd{k}\ .
\end{align*}

In this case, as expected, we find each contribution to vanish. The contribution of the linear term in $m$ can easily be shown to vanish. In fact, if we expand
\[
  M^{(d)}\rnd{k-\frac{d+1}{2}} = \sum_{r=0}^{d-1} a_r k^r\ ,
\]
then
\[
  (-)^{\frac{d}{2}}\Delta c_1 = - 4 m \sum_{r=0}^{d-1} a_r \zeta(-r)\ ,
\]
and
\begin{align*}
  (-)^{\frac{d}{2}}\Delta c_2 = & 4 R \frac{\partial}{\partial R} m \sum_{k=0}^{\infty} \sum_{r=0}^{d-1} a_r k^r  \psi\rnd{k}\\
  & = 4 R \frac{\partial}{\partial R} m \sum_{k=0}^{\infty} \sum_{r=0}^{d-1} a_r k^r \sqr{log(k) +
   \sum_{s=0}^{\infty} \zeta(-s) k^{-s-1}}\\
   & = 4 m  \sum_{r=0}^{d-1} a_r \zeta(-r)\ .
\end{align*}

The coefficients of the higher powers of $m$ also turn out to be zero, due to nontrivial cancellations in the expansion in $k$. It may be possible to give a general argument that shows this is true in any dimension, but we think this would add little understanding, and we were content of checking that the coefficients vanish for the first few dimensions.

\section{Conclusion}

Let us summarize the content of this paper
\begin{itemize}
  \item We have discussed how to impose the boundary conditions on the fields of the gravity theory. We have shown that the appropriate boundary conditions arise in a very natural way from the variational principle, if the appropriate boundary terms are added to the action. In this framework, the gravity action that is dual to a pure conformal field theory is not just the naive action, but needs to be complemented by an holographic renormalization boundary term. This term, besides being important for having a finite on-shell action, is also important in the determination of the boundary conditions. Perturbations of the conformal field theory like double trace deformations or source terms can then be introduced by adding additional boundary terms, whose form exactly matches the form of the CFT perturbation.
  \item We have reviewed the $c$-conjecture, both Cardy's proposal and the recent proposal in \cite{Myers:2010xs} of a $c$-function from entanglement entropy that is meaningful in any dimensions. In even dimensions, when both $c$-functions can constructed, we have shown that they coincide, at least from their holographic construction. 
  \item We have verified that, for this proposal of $c$ function, the $c$-conjecture holds along the RG flow induced by a double trace deformation, at least in the sense that the difference in central charge between the end point and the start point of the flow is negative. This result was already known for the case of a double trace deformation by a scalar operator. We have shown that it also holds for a spinor operator, and we have computed the difference in central charge from both holographic and CFT methods, finding that the two results agree.
\end{itemize}

\vspace{4 mm}
{\noindent \Large \bf Acknowledgments}
\vspace{4 mm}

Work supported in part by funds provided by the U.S. Department of Energy 
(D.O.E.) under cooperative research agreement DE-FG0205ER41360. We would like to thank John McGreevy for his help through many useful discussions and Igor Klebanov for important pieces of advice.

\section{Appendix}
\subsection{Computation of the propagator for the scalar field on AdS space}

The propagator is defined by
\[
  \sqrt{g}\rnd{-g_{\mu\nu} D_\mu D_\nu + \frac{m^2}{L^2}}G(r,x;s,y) = \delta(r-s)\delta^d(x-y)\ ,
\]
or, more explicitly,
\[
  \frac{L^{d-1}}{r^{d+1}}\sqr{-r^2\partial_r^2 + (d-1) r \partial_r - r^2\nabla^2 + m^2}G(r,x;s,y) = \delta(r-s)\delta^d(x-y)\ .
\]

Rescaling the propagator as
\[
  G(r,x;s,y) = \frac{r^{\frac{d}{2}}s^{\frac{d}{2}}}{L^{d-1}} D(r,x;s,y)\ ,
\]
and expanding in plane waves in the transverse directions
\[
  D(r,x;s,y) = \int \frac{d^d k}{(2\pi)^d} e^{i k\cdot(x-y)} D(r,s;k)\ ,
\]
the equation becomes
\[
  \rnd{-\partial_r^2 -\frac{1}{r} \partial_r +\frac{\nu^2}{r^2} + k^2}D(r,s;k) = \frac{1}{s}\delta(r-s)\ .
\]

Now we want to expand in eigenfunctions of the differential operator, and we have
\[
  \rnd{-\partial_r^2 -\frac{1}{r} \partial_r +\frac{\nu^2}{r^2}}J_{\pm \nu}(\omega r) = \omega^2 J_{\pm \nu}(\omega r)\ .
\]

Now the question of the boundary conditions on fields becomes important, in order to choose the sign on $\nu$. If we use the $\Delta_+$ action, then the field must go to zero approaching the boundary. Since $J_{-\nu}(r)$ diverges as $r \to 0$, this implies that we have to expand over $J_\nu$. If we use the $\Delta_-$ action, then we want
\[
  \pi = \sqr{r \partial_r  - \frac{d}{2} + \nu} \phi
\]
to go to zero. Taking into account the fact that we rescaled the fields, it is not difficult to see that this forces us to expand over $J_{-\nu}$. This means that we can just compute for the $\Delta_+$ case, the $\Delta_-$ case will be obtained by simply sending $\nu \to -\nu$.

Then let us then expand\footnote{The extra $\omega$ factor is the weight with respect to which the Bessel functions are orthonormal} over $J_\nu$:
\[
  D(r,s;k) = \int \dd \omega\ \omega\ J_\nu(\omega r) D(w,s;k)\ ,
\]
and, using the completeness relation
\[
  \int \dd \omega\ \omega\ J_\nu(\omega r) J_\nu(\omega s) = \frac{1}{s} \delta(r-s)\ ,
\]
we have
\[
  (\omega^2 + k^2) D(\omega,s;k) = J_\nu(\omega s)\ ,
\]
that is
\[
  D(r,s;k) = \int \dd \omega\ \omega\ \frac{J_\nu(\omega r)J_\nu(\omega s)}{\omega^2 + k^2}\ ,
\]
and hence
\[
  G(r,x;s,y) = \frac{r^{\frac{d}{2}}s^{\frac{d}{2}}}{L^{d-1}}\int \frac{d^d k}{(2\pi)^d} e^{i k\cdot(x-y)} \int \dd \omega\ \omega\ \frac{J_\nu(\omega r)J_\nu(\omega s)}{\omega^2 + k^2}\ .
\]

\subsection{Computation of the propagator for the spinor field on AdS space}

The propagator is defined by
\[
  \sqrt{g}\rnd{\ovslash{D} - \frac{m}{L}} G(r,x; s,y) = \delta(r-s) \delta(x - y)\ ,
\]
or, more explicitly
\[
  \frac{L^{d}}{r^d}\sqr{\gamma^r\rnd{\partial_r - \frac{d}{2 r}} + \gamma \cdot \partial - \frac{m}{r}} G(r,x s,y) = \delta(r-s) \delta(x - y)\ .
\]

Let us now rescale the propagator and expand in plane waves
\[
  G(r,x;s,y) = \frac{r^{\frac{d+1}{2}}s^{\frac{d+1}{2}}}{L^{d}} D(r,x;s,y)\ ,
\]
\[
  D(r,x;s,y) = \int \frac{\dd^d k}{(2\pi)^d} e^{ik \cdot(x-y)}D(r,s;k)\ .
\]

Then we have
\[
\sqr{\gamma^r\rnd{\partial_r + \frac{1}{2 r}} + i \gamma \cdot k - \frac{m}{r}} D(r,s; k) = \frac{1}{s}\delta(r-s)\ .
\]

Now let us consider all the projections of this equation over the eigenspaces of $\gamma^r$. We have
\[
\left\{
\begin{array}{l}
  \displaystyle
  \pm \rnd{\partial_r + \frac{1/2\mp m}{r}} D_{\pm\pm} + i \gamma \cdot k D_{\mp\pm} = P_\pm \frac{1}{s}\delta(r-s)\\ \\
  \displaystyle
  \pm\rnd{\partial_r + \frac{1/2\mp m}{r}} D_{\pm\mp} + i \gamma \cdot k D_{\mp\mp} = 0\ ,
\end{array}
\right.
\]
where
\[
  D(r,s;k) = P_\alpha D_{\alpha\beta}(r,s;k)P_\beta\ ,
  \quad \alpha, \beta \in \{+,-\}\ ,
\]
and
\[
  P_{\pm} = \frac{1\pm \gamma^r}{2}\ .
\]

Solving the homogeneous equations for $D_{-+}$ and $D_{+-}$ and substituting in the inhomogeneous we get
\[
  \sqr{\partial_r^2 + \frac{1}{r} \partial_r - \frac{(m+1/2)^2}{r^2} - k^2} D_{-+}(r,s;k) = P_-\ i \gamma \cdot k\ P_+\ \frac{1}{s}\delta(r-s)\ ,
\]
\[
  \sqr{\partial_r^2 + \frac{1}{r} \partial_r - \frac{(m-1/2)^2}{r^2} - k^2} D_{+-}(r,s;k) = P_+\ i \gamma \cdot k\ P_-\ \frac{1}{s}\delta(r-s)\ .
\]

The differential operator on the left hand side is hermitian, so we would like to expand in its eigenfunctions. In fact we have
\[
  \sqr{\partial_r^2 + \frac{1}{r} \partial_r - \frac{\nu^2}{r^2}}J_{\pm\nu}(\omega r) = - \omega^2 J_{\pm \nu}(\omega r)\ ,
\]
and the Bessel functions $J_{\nu}(\omega r)$ are orthonormal and complete
\[
  \int_{0}^{\infty} \dd \omega\ \omega\ J_{\nu}(\omega r) J_{\nu}(\omega s) = \frac{1}{s} \delta(r-s)\ .
\]

At this point boundary conditions become important, in order to understand what sign of $\nu$ to choose for each equation. Looking again at the variation of the $\Delta_+$ action we have:
\[
  \delta S_+ = \mathrm{e.o.m.} + \int \dd^d x \bar \psi P_+ \delta\psi + \delta \bar \psi P_- \psi\ .
\]

So we have to impose the following boundary conditions
\[
  P_- \psi = 0, \quad
  \bar \psi P_+ = 0\ ,
\]
and this leads to the requirement
\[
  \lim_{r\to 0} D_{-+}(r,s;k) = \lim_{s\to 0} D_{-+}(r,s;k)= 0\ .
\]

Therefore we have to use the expansion
\[
  D_{-+}(r,s;k) = \int_0^\infty \dd \omega\ \omega\ D_{-+}(\omega,s;k) J_{m+\frac{1}{2}}(\omega r)\ ,
\]
whereas the opposite choice of using $J_{-m -\frac{1}{2}}$ would give a divergent $r \to 0$ limit. 

On the other hand, there are no conditions to be imposed on $D_{+-}$, and both expansions can be used. We choose to expand over $J_{m-\frac{1}{2}}$, because it leads to slightly more symmetric expressions, but the same results could be obtained by expanding over $J_{-m+\frac{1}{2}}$
\[
  D_{+-}(r,s;k) = \int_0^\infty \dd \omega\ \omega\ D_{+-}(\omega,s;k) J_{m-\frac{1}{2}}(\omega r)\ .
\]

Substituting the expansion in the equations we get
\[
  D_{-+}(r,s;k) = - \int_0^\infty \dd \omega\ \omega\ \frac{P_-\ i\gamma \cdot k\ P_+}{\omega^2 + k^2} J_{m+\frac{1}{2}}(\omega r)J_{m+\frac{1}{2}}(\omega s)\ ,
\]
\[
  D_{+-}(r,s;k) = - \int_0^\infty \dd \omega\ \omega\ \frac{P_+\ i\gamma \cdot k\ P_-}{\omega^2 + k^2} J_{m-\frac{1}{2}}(\omega r)J_{m-\frac{1}{2}}(\omega s)\ .
\]

Now we can substitute in the homogeneous equations and get an expression for $D_{++}$ and $D_{--}$ as well. The final form of the propagator is
\begin{align*}
  & G(r,x;s,y) = - \frac{r^{\frac{d+1}{2}}s^{\frac{d+1}{2}}}{L^{d}}\int \frac{\dd^d k}{(2\pi)^d} e^{ik \cdot(x-y)} \int_0^\infty \dd \omega\ \omega\ \times\\
  &\times \sqr{ J_{m-\frac12}(\omega r) P_+ + J_{m+\frac12}(\omega r) P_-} \frac{\omega + i\gamma \cdot k }{k^2 + \omega^2}\sqr{P_+ J_{m+\frac12}(\omega s) + P_- J_{m-\frac12}(\omega s)}\ .
\end{align*}

\subsection{Computation of the spectrum of the spinor two point function on the sphere}

Since the theory is supposed to be Weyl invariant, i.e. invariant upon the simultaneous transformation
\begin{align*}
  &h(x) \to h'(x) = e^{\omega(x)} h(x) \\
  &\chi(x) \to \chi'(x) = e^{-\Delta\omega(x)} \chi(x)\\
  &\bar \chi(x) \to \bar \chi'(x) =  e^{-\Delta\omega(x)} \bar\chi(x)\ ,
\end{align*} 
we can obtain an expression for the two point function
\[
  G(\theta,\theta') = \exval{\chi(\theta) \bar \chi(\theta')}_{\scriptscriptstyle\mathrm{CFT}}
\]
on the sphere by operating a Weyl transformation on the flat space result. Consider the flat metric in polar coordinates
\[
  h_{E^d} = \dd r^2 + r^2 \dd \Omega_{d-1}^2\ ,
\]

Using the stereographic projection from the south pole, $r = R \tan \frac{\theta_d}{2}$, we can show that the flat metric is conformally equivalent to the metric on the sphere
\[
  h_{E^d} = \frac{1}{\rnd{2 \cos^2 \frac{\theta}{2}}^2} R^2\rnd{\dd \theta_d^2 + \sin^2 \theta_d \dd \Omega_{d-1}^2} = \frac{1}{\rnd{2 \cos^2 \frac{\theta}{2}}^2}\ h_{S^d}\ .
\]

Up to an irrelevant overall normalization factor, the correlation function in flat space is
\[
  \exval{ \chi(x) \bar \chi(0)}_{E^d} = \frac{\gamma\cdot x}{|x|^{2\Delta + 1}}\ .
\]

Going to polar coordinates and then to the projective coordinates we have
\[
  \exval{ \chi(\theta_d,\Omega_{d-1}) \bar \chi(0,0)}_{E^d} = \frac{\gamma^d\cos \theta_{d-1} + \gamma^{d-1}\sin\theta_{d-1}\cos \theta_{d-2}+ \ldots}{\rnd{R\tan\frac{\theta_d}{ 2}}^{2\Delta}}\ ,
\]
and, performing the Weyl rescaling, we can obtain the correlator on the sphere
\[
  \exval{ \chi(\theta_d,\Omega_{d-1}) \bar \chi(0,0)}_{S^d} = \frac{\gamma^d\cos \theta_{d-1} + \gamma^{d-1}\sin\theta_{d-1}\cos \theta_{d-2}+ \ldots}{\rnd{2 
  R\sin\frac{\theta_d}{ 2}}^{2\Delta}}\ .
\]

In particular we will need the result on the principal meridian $\Omega_{d-1} = 0$
\[
  G(\theta) = \frac{\gamma^d}{\rnd{2 
  R\sin\frac\theta 2}^{2\Delta}}\ .
\]

We assume that $G$ is diagonal if expanded over the eigenfunctions of the Dirac operator $\ovslash{D}^{(d)}$ on the sphere\footnote{For the moment we will work on a sphere of unit radius}
\[
  \ovslash{D}^{(d)} = \gamma^d\rnd{\frac{\partial}{\partial\theta_d} + \frac{d-1}{2} \cot \theta_d} + \frac{1}{\sin \theta _d} \ovslash{D}^{(d-1)};
  \quad
  \ovslash{D}^{(0)} = 0\ .
\]

The eigenfunctions are known \cite{Camporesi:1995fb},  let us quickly review their properties. They are labeled by two set of quantum numbers
\[
  n = \{n_1,\ldots,n_d\}\ , \quad
  0\leq n_1\leq n_2\leq \ldots \leq n_d\ ,
\]
\[
  \sigma = \{\sigma_1, \ldots, \sigma_d\}\ ,\quad
  \sigma_i \in \{-1,1\}\ ,
\]
and they satisfy
\[
  \ovslash{D}^{(d)} \psi^{(d)}_{\sigma n}(\Omega_d) = i \sigma_d \rnd{n_d + \frac{d}{2}}\psi^{(d)}_{\sigma n}(\Omega_{d})\ .
\]

They can be constructed by recursion:
\begin{align*}
  \psi^{(d)}_{\sigma n}(\theta_d,\Omega_{d-1}) = & c^{(d)}_{n_d n_{d-1}}\Big[\phi^{(d)}_{\sigma_{d-1} n_d n_{d-1}}(\theta_d)(1+i\gamma^d) +\\
  & + \sigma_d \sigma_{d-1} \phi^{(d)}_{-\sigma_{d-1} n_d n_{d-1}}(\theta_d) (1-i\gamma^d)\Big]\psi^{(d-1)}_{\sigma n}(\Omega_{d-1})\ ,
\end{align*}
\[
  \psi^{(1)}_{\sigma n}(\theta_1) = \frac{1}{\sqrt{2\pi s}}e^{i \sigma_1\rnd{n_1+\frac12}\theta_1}
  \begin{pmatrix}1\\ \vdots\\1\end{pmatrix}\ , \quad 
  s = \dim \gamma \ ,
\]
where
\[
  \phi^{(d)}_{+ n l}(\theta) = \rnd{\cos\frac{\theta}{2}}^l \rnd{\sin\frac{\theta}{2}}^{l+1}P_{n-l}^{(\frac{d}{2}+l,\frac{d}{2}+l-1)}(\cos\theta)\ ,
\]
\[
  \phi^{(d)}_{- n l}(\theta) = \rnd{\cos\frac{\theta}{2}}^{l+1} \rnd{\sin\frac{\theta}{2}}^{l}P_{n-l}^{(\frac{d}{2}+l-1,\frac{d}{2}+l)}(\cos\theta)\ ,
\]
and where $P_n^{(a,b)}$ are the Jacobi polynomials. With this construction not all values of the quantum numbers $\sigma$ label different eigenfunctions. In fact, for $i<d/2$, $\sigma_{2i} = \pm 1$ labels the same eigenfunction. This has to be kept in mind when computing the degeneracy of an energy level.

The normalization is fixed by requiring
\[
  \int \dd \Omega_d \psi_{\sigma n}^{(d)\dag}(\Omega_d) \psi^{(d)}_{\sigma n}(\Omega_d) = 1\ ,
\]
and we have
\[
  c^{(d)}_{n l} = \frac{\sqrt{\Gamma(n-l+1)\Gamma(d+n+l)}}{2^{\frac{d}{2}}\Gamma\rnd{\frac{d}{2}+n}}\ .
\]

Now we can write explicitly the expansion of $G$ over the eigenfunctions:
\[
  G(\Omega_d,\Omega'_d) = \sum_{n, \sigma} (-i) \sigma_d g_{n_d} R^{-d}\psi_{\sigma n}^{(d)}(\Omega_d) \psi^{(d)\dag}_{\sigma n}(\Omega'_d)\ ,
\]
where the factor $R^{-d}$ is necessary because the eigenfunctions $\psi_{\sigma n}^{(d)}$ are normalized on the unit sphere.

In order to get a simple expression,  we let $\theta'_d = 0$, $\Omega'_{d-1} = \Omega_{d-1} = 0$, i.e. we take the second point to be the north pole of the sphere and the first point to lie on the principal meridian. In this case, with some tedious work, it can be shown that
\begin{align*}
  \sum_{\sigma} \sigma_d \psi_{n\sigma}^{(d)}(\theta,0) & \psi^{(d)\dag}_{n\sigma}(0,0) = \ \delta_{n_{d-1}0}\delta_{n_{d-2}0}\ldots\delta_{n_{1}0}\ \frac{i \gamma^d}{\Sigma_{d-1}} \times \\
  &\times 4 \rnd{c^{(d)}_{n_d0}}^2 \sin\frac{\theta}{2}\ P_{n_d}^{(\frac{d}{2},\frac{d}{2}-1)}(\cos\theta)\ P_{n_d}^{(\frac{d}{2}-1,\frac{d}{2})}(1)\ ,
\end{align*}
where $\Sigma_{d}$ is the volume of the $d$-sphere of unit radius, and hence
\begin{align*}
  G(\theta) & \equiv G(\theta,0; 0, 0) \\
   &= \sum_{n=0}^{\infty} g_n R^{-d}\frac{\gamma^d}{\Sigma_{d-1}} 4 \rnd{c^{(d)}_{n0}}^2 \sin\frac{\theta}{2}\ P_{n}^{(\frac{d}{2},\frac{d}{2}-1)}(\cos\theta)\ P_{n}^{(\frac{d}{2}-1,\frac{d}{2})}(1)\ .
\end{align*}

Now we can extract the eigenvalues $g_n$ by exploiting the orthogonality relation of the Jacobi polynomials
\begin{align*}
  \int_{-1}^{1} \dd x\ &(1-x)^a (1+x)^b P_n^{(a,b)}(x)P_m^{(a,b)}(x)=\delta_{mn} \times\\ 
  &\times\frac{2^{a+b+1}}{2n + a + b +1} \frac{\Gamma(n+a+1) \Gamma(n+b+1)}{\Gamma(n+1) \Gamma(n+a+b+1)}\ ,
\end{align*}
so that we have
\[
  g_n = \frac{\Sigma_{d-1}R^{d}}{P_n^{(\frac{d}{2}-1,\frac{d}{2})}(1)} \int_{-1}^{1} \dd x\ (1-x^2)^{\frac{d}{2}-1} \sqrt{\frac{1-x}{2}} \frac{G(x)}{\gamma^d} P_n^{(\frac{d}{2},\frac{d}{2}-1)}(x)\ ,
\]
where $x = \cos \theta$ and by $G(x)/\gamma^d$ we mean the coefficient that multiplies $\gamma^d$ in $G$.

Explicitly
\[
  g_n \propto \frac{R^{d-2\Delta}}{P_n^{(\frac{d}{2}-1,\frac{d}{2})}(1)} \int_{-1}^{1} \dd x\ (1-x^2)^{\frac{d}{2}-1} (1-x)^{\frac{1}{2}-\Delta} P_n^{(\frac{d}{2},\frac{d}{2}-1)}(x)\ ,
\]
where we have dropped an overall prefactor that does not depend on $n$ or $R$. Such factor only amounts to a change in the normalization of the operator and can be ignored. Or, from an alternative point of view, when computing the central charge, this factor would produce an additive constant that would not survive the derivative with respect to $R$. The integral can be done exactly, and the $n$- and $R$ -dependent part of the result is
\[
  g_n \propto R^{d-2\Delta}\frac{\Gamma\rnd{n+\Delta +\frac{1}{2}}}{\Gamma\rnd{n+d-\Delta +\frac{1}{2}}}\ .
\]

The degeneracy of each eigenvalue can be obtained from the constraints on the quantum numbers of the eigenfunctions:
\[
  M^{(d)}(n) = \dim\gamma \frac{(n+d-1)!}{n!(d-1)!}\ .
\]

\bibliographystyle{unsrt}
\bibliography{/home/andrea/Include/latex/holography.bib}

\end{document}